\documentclass{article}

\usepackage{PRIMEarxiv}
\usepackage{makecell}
\usepackage{tabularx}
\usepackage[utf8]{inputenc} 
\usepackage[T1]{fontenc}    
\usepackage{orcidlink}
\usepackage{hyperref}
\hypersetup{hidelinks}
\usepackage{url}            
\usepackage{booktabs}       
\usepackage{amsfonts}       
\usepackage{nicefrac}       
\usepackage{array}
\usepackage{microtype}      
\usepackage{lipsum}
\usepackage{fancyhdr}       
\usepackage{graphicx}       
\graphicspath{{}}     
\usepackage{textcomp} 
\usepackage[acronym]{glossaries}
\makeglossaries
\newacronym{oss}{OSS}{Open source software}
\newacronym{ngi}{NGI}{Next Generation Internet}
\newacronym{sta}{STA}{Sovereign Tech Agency}
\newacronym{stf}{STF}{Sovereign Tech Fund}
\newacronym{chaoss}{CHAOSS}{Community Health Analytics Open Source Software}
\newacronym{lf}{LF}{Linux Foundation}
\newacronym{eu}{EU}{European Union}
\newacronym{ec}{EC}{European Commission}
\newacronym{roi}{ROI}{return on investment}
\newacronym{fstp}{FSTP}{Financial Support for Third Parties}
\newacronym{gdp}{GDP}{gross domestic product}
\newacronym{ria}{RIA}{Research and Innovation Action}
\newacronym{rias}{RIAs}{Research and Innovation Actions}
\newacronym{csa}{CSA}{Coordination and Support Action}
\newacronym{csas}{CSAs}{Coordination and Support Actions}
\newacronym{loc}{LOC}{lines of code}

\title{A Toolkit for Measuring the Impacts of Public Funding on Open Source Software Development}

\author{
  Cailean Osborne\thanks{Corresponding author: Cailean Osborne, cailean.osborne@oii.ox.ac.uk} \orcidlink{0000-0002-4018-8488}\\
  University of Oxford \\
  Oxford, UK\\
 \AND
  Paul Sharratt \orcidlink{0009-0009-3459-3311} \\
  Sovereign Tech Agency \\
  Berlin, Germany\\
   \And
  Dawn Foster \orcidlink{0000-0002-8441-5873} \\
  The CHAOSS Project \\
  London, UK \\
 \AND
  Mirko Boehm \orcidlink{0000-0001-7658-3896}  \\
  The Linux Foundation \\
  Berlin, Germany\\
}

\begin{document}
\maketitle
\vspace{-1em}

\begin{abstract}
Governments are increasingly employing funding for open source software (OSS) development as a policy lever to support the security of software supply chains, digital sovereignty, economic growth, and national competitiveness in science and innovation, among others. However, the impacts of public funding on OSS development remain poorly understood, with a lack of consensus on how to meaningfully measure them. This gap hampers assessments of the return on public investment and impedes the optimisation of public-interest funding strategies. We address this gap with a toolkit of methodological considerations that may inform such measurements, drawing on prior work on OSS valuations and community health metrics by the Community Health Analytics Open Source Software (CHAOSS) project  as well as our first-hand learnings as practitioners tasked with evaluating funding programmes by the Next Generation Internet initiative and the Sovereign Tech Agency. We discuss salient considerations, including the importance of accounting for funding objectives, project life stage and social structure, and regional and organisational cost factors. Next, we present a taxonomy of potential social, economic, and technological impacts that can be both positive and negative, direct and indirect, internal (i.e. within a project) and external (i.e. among a project's ecosystem of dependents and users), and manifest over various time horizons. Furthermore, we discuss the merits and limitations of qualitative, quantitative, and mixed-methods approaches, as well as options for and hazards of estimating multiplier effects. With this toolkit, we contribute the multi-stakeholder conversation about the value and impacts of funding on OSS developers and society at large.
\end{abstract}

\keywords{Open source software  \and funding \and sustainability \and impact measurement \and maintenance \and digital commons}

\section{Introduction}\label{sec:introduction}
\gls{oss}\footnote{\gls{oss} is software whose source code is distributed under a license that permits the use, study, modification, and redistribution of the software source code \cite{osi_open_2007}. Please note that we use \gls{oss} rather than FOSS, which stands for ``free and open source software''.} are digital public goods that are increasingly recognised as digital infrastructure \cite{eghbal_roads_2016,scott_avoiding_2023}, which are used in around 96\% of codebases \cite{synopsys_open_2023} and constitute up to 90\% of commercial software stacks \cite{open_source_security_foundation_oss_2022}.  The ubiquitous use of OSS, which are often developed and maintained by volunteer communities, have drawn attention to the question of funding as a mechanism to support the sustainability of OSS projects \cite{eghbal_working_2020,birkinbine_incorporating_2020}. In particular, the discovery of major security vulnerabilities in widely used OSS projects, such as the Log4Shell vulnerability in November 2021, highlighted the need to fund security and maintenance work in critical OSS projects \cite{vaughan-nichols_log4shell_2021}, and contributed to an understanding of the role and responsibility of the public sector as a funder of OSS development  \cite{herpig_fostering_2023, keller_european_2022}. Concurrently,  policymakers increasingly recognise OSS funding as a policy lever to support digital sovereignty \cite{osborne_european_2023,burwell_digital_2022}, the growth of domestic software markets \cite{nagle_government_2019}, and national competitiveness in science and innovation \cite{osborne_public-private_2024,bloom_toolkit_2019}, among others. 

While governmental interest and involvement in funding OSS funding is increasing, its impacts on OSS development is poorly understood, with no consensus on how to measure them in a meaningful way. Not only is the measurement of the impacts of OSS funding methodologically challenging, but it is complicated by the fact that introducing funding into OSS projects may change contributor incentives and the balance of voluntary and paid participation \cite{bohm_economics_2022, gerosa_shifting_2021}. This gap hinders assessments of return on public investment and optimising public interest funding strategies. We address this problem with a toolkit of methodological considerations for measuring the impacts of public funding on OSS development. It is informed by methodologies for OSS valuations and community health measurements, in particular metrics developed by the \gls{chaoss} project, as well as our first-hand insights from developing impact measurement frameworks for the \gls{ngi} initiative at the \acrfull{ec} and the \gls{sta} in Germany. 

The toolkit begins with a discussion of key considerations, including funding objectives, project life stages and social structures, and regional and organisational cost factors. Then, we present a taxonomy of potential social, economic, and technological impacts that can be both positive and negative, direct and indirect, internal (i.e. within a project) and external (i.e. among a project's ecosystem of dependents and users), and manifest over various time horizons. Next, we discuss the merits and limitations of qualitative, quantitative, and mixed-methods approaches for measuring such impacts, and options for and hazards of estimating multiplier effects. This toolkit is not exhaustive or prescriptive, nor does it aim to be; rather, we seek to provide a toolkit that can inform the multi-stakeholder debate about the value of public funding for OSS development and how to meaningfully measure its impacts.

This paper has the following structure. Section~\ref{sec:lit-review} reviews relevant literature and practice, providing a background on prior scholarship on OSS funding, OSS valuations, and OSS community health measurements, as well as real-world examples of public funding by the \gls{ngi} and \gls{sta}. Section~\ref{sec:toolkit} presents the toolkit for measuring the impacts of public funding on OSS development. It includes a discussion of pertinent considerations, such as funding objectives (\ref{subsubsec:toolkit-objectives}); salary structures (\ref{subsubsec:toolkit-costs}); a taxonomy of social, economic, and technological impacts (\ref{subsec:toolkit-taxonomy}); multiplier effect estimations (\ref{subsec:toolkit-multiplier-effects}); and an evaluation of qualitative, quantitative, and mixed-methods approaches (Section~\ref{subsec:toolkit-methodological-approaches}). Section~\ref{sec:discussion} discusses the overarching considerations and future research directions. Finally, Section~\ref{sec:conclusion} concludes with a call to action for both diverse OSS stakeholders to engage in the debate about the value and impacts of public funding on OSS development.

\section{Review of Literature \& Practice}\label{sec:lit-review}
\subsection{Review of Prior Work}\label{subsec:prior-work}
\subsubsection{Funding OSS Development: History, Funders, and Impacts }
OSS are digital public goods \cite{dulong_de_rosnay_digital_2020} that are increasingly recognised as digital infrastructure \cite{eghbal_roads_2016,scott_avoiding_2023}, owing to their use in around 96\% of codebases \cite{synopsys_open_2023} and constitution of up to 90\% of commercial software stacks \cite{open_source_security_foundation_oss_2022}. In light of their ubiquitous use, the topic of funding for development, maintenance, and security of OSS has become increasingly important in developer, industry, and policy circles alike. In particular, high-profile security vulnerabilities like the Heartbleed bug in 2014, the Log4Shell vulnerability in 2021, or the recent XZ Utils backdoor in 2024 have underscored the unsustainability of relying on volunteer communities that maintain a large portion of critical OSS projects \cite{vaughan-nichols_log4shell_2021}. 

Historically, money has been a contentious issue among OSS developers, many of whom take pride in the origins of OSS as a social movement against the privatisation of software or as a ``programmers' paradise'' \cite{raymond_cathedral_2001}, and who do not contribute for financial gain \cite{gerosa_shifting_2021}, but out of their intrinsic satisfaction of learning and solving technical problems \cite{loebbecke_open_2003}, their altruism \cite{markus_what_2000}, or their political ideals \cite{kelty_two_2008}, among others. Some even argue that ``money ruins everything'' \cite{eghbal_where_2017}. In addition, evidence from the Rust community illustrates that volunteers have some prejudices against paid developers, such as that they do ``do boring work'', ``rarely care [for] documentation'', and ``lack personal attachment'' \cite{zhang_how_2024}.\footnote{This study also finds differences in the activity between volunteer and paid developers; for example, paid developers tend to contribute more frequently and implement more features than volunteers \cite{zhang_how_2024}.} However, growing concerns about maintainer burnout \cite{eghbal_working_2020}, security vulnerabilities \cite{vaughan-nichols_log4shell_2021}, and widespread commercial freeriding on the labour of volunteers \cite{geiger_labor_2021,birkinbine_incorporating_2020}
have changed minds in OSS developer communities \cite{osborne_public-private_2024,salkever_open_2023}. As Benjamin Birkinbine contends, there is an urgent need ``not just [for] investment in institutions, organisations, technologies, or innovations, but long-term and sustainable investment in the true source of their value: people'' \cite{birkinbine_incorporating_2020}. \newline 

\textit{\textbf{Micro-donations: Overview and Impacts}}\newline

Various actors have been involved in funding OSS development, including individuals, companies, philanthropic organisations, governments, and public research agencies. Individuals, typically users and/or developers of OSS, contribute through micro-donations via platforms like NumFOCUS and GitHub Sponsors \cite{github_sponsoring_2023}. Motivations for micro-donations include individuals' desire to show their appreciation for an OSS project and to encourage further development \cite{zhang_who_2021}. While the relative impact of micro-donations appears limited \cite{overney_how_2020}, they have been shown to shorten maintainer response times to issue reports \cite{nakasai_analysis_2017,nakasai_are_2019}, increase maintenance-related activities \cite{medappa_sponsorship_2023}, and facilitate project spending on community events and summer internships for developers from under-represented groups \cite{osborne_public-private_2024}. \newline

\textit{\textbf{Commercial Funding: Overview and Impacts}}\newline

The private sector has been the largest direct and indirect funder of OSS development to date, which is unusual among public goods. It is common for companies to sponsor OSS developers and to let their employees contribute to OSS projects during work hours, either as a job responsibility or as part of a voluntary initiative \cite{butler_investigation_2018,dahlander_man_2006,xia_lessons_2023}. Companies also sponsor OSS foundations and consortia, paying membership fees to the organisations that host projects \cite{omahony_boundary_2008}. The impact of such funding is contested: while only 12\% of a Tidelift survey respondents found this type of funding very useful \cite{tidelift_2020_2020}, a case study on the \textit{scikit-learn} project found that commercial sponsorship of its consortium was useful for balancing the influence of companies and provided resources to employ full-time maintainers \cite{osborne_public-private_2024}. Some companies run OSS contribution programmes, such as the Google Summer of Code programme \cite{hawthorn_google_2008}, and some have established ``FOSS Contributor Funds'' that allow employees to nominate OSS projects that they would like to sponsor \cite{obrien_foss_2019,obrien_sustaining_2019}. However, these funds typically offer limited amounts for short periods of time (e.g. \$10,000 for one year).  In addition, companies spin-out proprietary software into OSS projects and invest in their ecosystems as a strategy to increase adoption of their software, benefit from external contributions, or reduce a competitor’s market share \cite{west_contrasting_2005,osborne_characterising_2024}. 

The economic value of commercial funding and investments in OSS development are significant by many measures. It has been estimated that €1 billion invested in OSS by companies in the \gls{eu} in 2018 generated up to €95 billion for the EU's \gls{gdp} that year \cite{blind_impact_2021}, and that companies in the USA invested \$37.8 billion in OSS in 2019 \cite{korkmaz_github_2024}. Furthermore, country-level OSS development activity and the founding of start-ups are positively correlated, suggesting that OSS development activity is a catalyst for innovation \cite{wright_open_2023}.  Nevertheless, we face a stark imbalance between the supply-side value (\$4.15 billion) of OSS development and its demand-side value (\$8.8 trillion)\cite{hoffmann_value_2024}. These numbers underline the ``risk of underproduction'' stemming from the misalignment of the supply of OSS development labour and the demand for OSS \cite{champion_underproduction_2021}. Beyond funding, the impacts of commercial participation on collaborative practices and norms in OSS developer communities is an active area of research \cite{li_systematic_2024,germonprez_eight_2018,osborne_why_2024}. \newline

\textit{\textbf{Public Funding: Overview and Impacts}}\newline

While the public sector is often overlooked as a major funder,  it has funded OSS directly and indirectly since the inception of the world wide web at CERN through research grants \cite{osborne_public-private_2024,bloom_toolkit_2019}.  The intensity of public funding for OSS development has changed in recent years, with the recognition of funding as a policy instrument to enhance the security of software supply chains and digital infrastructure \cite{herpig_fostering_2023},  digital sovereignty of nations and blocs \cite{osborne_european_2023,burwell_digital_2022}, and national competitiveness in science and innovation \cite{osborne_public-private_2024}.\footnote{We note that beyond OSS direct funding for R\&D is recognised as one of the most effective, short-term policy instruments for promoting innovation \cite{bloom_toolkit_2019}.}  Examples of public funding include research grants for innovation \cite{osborne_public-private_2024}; award and prize programmes for OSS developers, such as the BlueHat Prizes by France's Free Software Unit and NLNet \cite{guerry_codegouvfr_2024}; bespoke OSS funding bodies, such as the Open Technology Fund in the USA, the \gls{ngi} initiative in the EU, and the \gls{sta} and Prototype Fund in Germany \cite{keller_european_2022}; and bug bounty programmes for security-focused maintenance of OSS projects \cite{ellis_opportunities_2024}.  However, the most common way in which governments have indirectly supported OSS to date has been through  procurement policies that favour OSS over proprietary solutions \cite{osborne_european_2023,lostri_government_2022}. Such policies have led to increased domestic OSS development activity and the growth of domestic software markets \cite{nagle_government_2019}. 

Prior work provides the following insights on the impact of public funding on OSS development. Few public research grants provide dedicated funding for scientific OSS development and those that do tend to prioritise innovation over maintenance \cite{strasser_ten_2022}. A case study on a €32 million grant awarded by the French government to the \textit{scikit-learn} project highlights both challenges and benefits of research grants, from its initial focus on innovation over maintenance and misalignments between policy objectives and the needs and know-how of the maintainers, to benefits for the project's long-term stability, ecosystem development, and maintenance capacity  \cite{osborne_public-private_2024}. Similarly, an evaluation of the ``Software Sprint'' funding programme by the Prototype Fund\footnote{The Prototype Fund is a funding program of the Federal Ministry of Education and Research (BMBF) that is managed and evaluated by the Open Knowledge Foundation Germany. Its ``Software Sprint'' distributed €12.3 million across 12 funding rounds to 293 projects that aimed to build public-interest OSS that addressed societal challenges.} shows mixed results. On the one hand, about two-thirds of funded projects successfully completed a functional prototype during their funding period and roughly 60\% continued development after funding ended \cite{galati_evaluation_2023}. In addition, the funding fostered the professional growth of individuals, with over a quarter developing new career opportunities. On the other hand, the programme faced challenges in supporting the sustainability of projects; for example, over half struggled to secure follow-up funding for scaling \cite{galati_evaluation_2023}. In addition, an evaluation of the \gls{ngi} initiative shows that it has supported the development of OSS projects that protect EU digital values and rights, enable EU legislation, and provide alternatives to proprietary solutions, among others \cite{european_commission_directorate-general_for_communications_networks_content_and_technology_benchmarking_2024}.

In addition, research illustrates the positive impact of funding for security work. There is a moderate positive correlation between general-purpose funding and enhanced security practices among the 1,000 most downloaded OSS packages in the Python and Javascript ecosystems \cite{meyers_o_2024}. In addition, the greater number of funding sources in a project corresponds with better security practices \cite{meyers_o_2024}. Bug bounty programmes enhance OSS security in mature projects, but should only be implemented after basic security practices are in place, as they can potentially overwhelm under-maintained projects and may disrupt the reciprocal nature of OSS communities unless properly designed \cite{ellis_opportunities_2024}. Holistic funding approaches that pair targeted security funding with general funding for maintenance are advised \cite{ellis_opportunities_2024}.

\subsubsection{OSS Valuation Models and Community Health Approaches}\label{subsubsec:chaoss}

An obstacle that stands in the way of funding for OSS development is that to make the argument to invest in OSS development, managerial decision-makers demand quantitative evidence of the value of OSS contributions \cite{vargas_fossy_2024}; for example, a \gls{roi} for a company or the responsible use of taxpayers' money for a public sector organisation. However, it is difficult 
to quantify the value of OSS contributions \cite{vargas_fossy_2024}. Motivated by the motto, ``If we can appropriately quantify the value of OSS work, then we can adequately invest in it,''  Sophia Vargas reviewed valuation models that can be used to measure the value of OSS  \cite{vargas_fossy_2024} (see Table~\ref{tab:vargas-valuation-models}). Vargas identifies several key challenges in estimating the value of OSS: the need for consistent measures of demand and usage \cite{vargas_estimating_2024}, the difficulty in quantifying non-code contributions \cite{young_which_2021}, and the ``force multiplier'' effect where collective benefits exceed individual inputs \cite{vargas_fossy_2024}. While economic models like \gls{gdp} may guide government funding strategies, methodological challenges (e.g. data availability and causality issues) limit the robustness of such measurements and they are less applicable to companies and individuals who are motivated by different types of value creation (e.g. skill development or ROIs). 

What is more, Vargas cautions against over-reliance on technical metrics like commits or \gls{loc}, which flatten socio-technical dynamics that characterise activity in OSS projects and their wider ecosystems of dependents and users. By contrast, socio-technical models that focus on the social structure and health of OSS projects provide a more comprehensive lens on value creation in OSS projects and ecosystem \cite{goggins_open_2021}. The health of an OSS project concerns ``an OSS project’s capability to stay viable and maintained over time without interruption or weakening'' \cite{linaker_how_2022}; or in other words, the sustainable availability of maintenance labour that can come from either the maintainers or the wider community \cite{linaker_sustaining_2024}. A key ingredient for the health of OSS projects is the provision of ``human infrastructure'' that can support and secure maintenance labour from both project maintainers and their community of contributors, including the securing of a work-life balance, the management of social pressure, and the diversity of contributors \cite{linaker_sustaining_2024}. The authors note funding as an important enabler for this human infrastructure \cite{linaker_sustaining_2024}.

While every OSS project is unique, Nadia Eghbal identifies four common social structures based on a project's contributor-to-user ratio: federations, clubs, stadiums, and toys \cite{eghbal_working_2020}. Federations have both high contributor and user growth, typically with complex governance processes and working groups. Clubs have high contributor growth but lower user growth, characterised by tight-knit communities of enthusiasts. Stadiums have high user growth but low contributor growth, typically maintained by a small group of developers. Toys are personal projects with low contributor and user growth. Beyond developers, Julia Ferraioli advocates for a social model of OSS that considers ``the people who consume, contribute to, and maintain it, rather than a technical model of open source, based on the legal rights and responsibilities as defined in the license'' \cite{ferraioli_bringing_2022}. Similarly, ``all contributors'' models consider both technical and non-technical contributions to OSS protects, providing a more comprehensive representation of contributorship \cite{young_which_2021}.  

Measuring the health of an OSS project is complex because there are many sociotechnical aspects to consider, from a project’s social composition (e.g. culture, governance, and contributor diversity) to its long-term stability (e.g. funding and code maintainability) \cite{linaker_how_2022}. For this reason, scholarship on community health and the metrics of the \gls{chaoss} project look beyond the code, considering social aspects like community welcomingness and diversity, equity, and inclusion; economic aspects, such as funding and business readiness; as well as technological aspects like security and license compliance \cite{chaoss_about_2023}. Furthermore, a key principle of the CHAOSS project is that every project is different, and accordingly metrics should always be interpreted with the needs and context of the project taken into account \cite{goggins_open_2021}. 


\begin{table}[htbp]
\centering
\caption{Valuation Models and their Application to OSS Development by Sophia Vargas (2024) \cite{vargas_estimating_2024}}
\label{tab:vargas-valuation-models}
\small
\setlength{\leftmargini}{1em}
\begin{tabular}{|p{4cm}|p{4cm}|p{4cm}|p{4cm}|}
\hline
\textbf{Model} & \textbf{Application to OSS} & \textbf{Key Variables} & \textbf{Challenges} \\
\hline
\vspace{1mm}
\textbf{Constructive Cost Model (COCOMO)}
\begin{itemize}
    \item COCOMO estimates the effort, cost, and schedule of software projects \cite{boehm_software_1981}
    \item COCOMO II includes reused/adapted code \& code incorporation effort \cite{boehm_cost_1995}
\end{itemize} &
\begin{itemize}
    \item \gls{loc} as proxy variable for value creation
    \item Complexity multipliers for context adjustment
\end{itemize} &
\begin{itemize}
    \item \gls{loc}
    \item Task complexity
\end{itemize} &
\begin{itemize}
    \item No maintenance tracking
    \item Does not account for code history or maintenance
\end{itemize} \\
\hline
\vspace{1mm}
\textbf{DevOps Research and Assessment (DORA)}
\begin{itemize}
    \item Evaluates maturity of DevOps performance and processes \cite{forsgren_accelerate_2018}
    \item Focus on operational metrics for DevOps evaluation \cite{puppet_2020_2020}
\end{itemize} &
\begin{itemize}
    \item Quality versus quantity measurement
    \item Compares upstream versus company-specific quality
\end{itemize} &
\begin{itemize}
    \item Deployment frequency
    \item Lead time for changes
    \item Change failure rate
    \item Time to restore service
\end{itemize} &
\begin{itemize}
    \item Assumes team structure
    \item Heterogeneity of OSS communities (e.g. varied incentives and time commitments)
\end{itemize} \\
\hline
\vspace{1mm}
\textbf{OSS as Socio-technical Model}
\begin{itemize}
    \item Complex software development processes by loosely coordinated developers \cite{scacchi_itr_2005}
    \item Social, economic, political network relationships \cite{kling_bit_2003}
    \item Social model: classify project by creator intent \cite{ferraioli_bringing_2022}
\end{itemize} &
\begin{itemize}
    \item Recognises OSS projects as complex systems
    \item Captures diversity of contributors and incentives
    \item Recognises multiple value types for individuals (skill development, learning, etc.)
\end{itemize} &
\begin{itemize}
    \item People (e.g. incentives)
    \item Infrastructure (e.g. tools)
    \item Content (e.g. documents)
    \item Governance (e.g. participation in roadmap decisions) 
\end{itemize} &
\begin{itemize}
    \item The more complexity one recognises, the more difficult it becomes to quantify value
    \item What kind of value, by whom, and for whom?
\end{itemize} \\
\hline
\vspace{1mm}
\textbf{Business Models (TCO/ROI)}
\begin{itemize}
    \item Total cost of ownership (TCO): a financial estimate of direct and indirect costs
    \item Return on investment (ROI): ratio between net income (over time) and investment 
\end{itemize} &
\begin{itemize}
    \item Business benefits of OSS \cite{chesbrough_measuring_2023}, some more quantifiable than others (e.g. cost savings vs. open standards creation)
    \item TCO includes infrastructure, maintenance, and integration
    \item \gls{roi} based on percentage of codebase that is OSS
\end{itemize} &
\begin{itemize}
    \item Infrastructure costs
    \item Maintenance expenses
    \item Community funding
    \item Integration effort
\end{itemize} &
\begin{itemize}
    \item Organisationally subjective
    \item Assumption of ``OSS is free''
    \item Difficult to measure \% of income associated with a OSS package or project
\end{itemize} \\
\hline
\vspace{1mm}
\textbf{Economic Models (e.g. GDP)}
\begin{itemize}
    \item \gls{gdp} measures market value of goods and services in a time period in a country
\end{itemize} &
\begin{itemize}
    \item Tracks economic impact
    \item Estimated: €1B invested in OSS in 2018 generated €65-95B for EU \gls{gdp} \cite{blind_impact_2021}
\end{itemize} &
\begin{itemize}
    \item OSS contributions
    \item Investments in OSS 
    \item Economic indicators
\end{itemize} &
\begin{itemize}
    \item Limited data availability
    \item Attribution difficulties
\end{itemize} \\
\hline
\vspace{1mm}
\textbf{Supply and Demand Models}
\begin{itemize}
    \item Market equilibrium analysis
\end{itemize} &
\begin{itemize}
    \item OSS creation versus replacement cost estimation \cite{hoffmann_value_2024}
    \item Estimated: Supply-side value = \$4.15 billion, demand-side value = \$8.8 trillion \cite{hoffmann_value_2024}.
\end{itemize} &
\begin{itemize}
    \item Value: Cost of software development labour (\gls{loc})
\end{itemize} &
\begin{itemize}
    \item Difficult to measure due to non-pecuniary nature and lack of usage tracking  \cite{hoffmann_value_2024}
    \item \gls{loc} focus ignores maintenance, non-technical labour, and different types of value
\end{itemize} \\
\hline
\vspace{1mm}
\textbf{Underproduction Risk Model}
\begin{itemize}
    \item Adapted supply and demand model for OSS context \cite{champion_underproduction_2021}
\end{itemize} &
\begin{itemize}
    \item Maintenance-focused
    \item Recognition of collaboration dynamics in OSS projects
\end{itemize} &
\begin{itemize}
    \item Maintenance supply
    \item Usage demand
\end{itemize} &
\begin{itemize}
    \item Correlational limitations
    \item Quality measurement issues
    \item Lack of usage data
\end{itemize} \\
\hline
\end{tabular}
\end{table}

\subsection{Review of Practice}\label{subsec:practice}
In this section, we review two public funding initiatives that we are familiar with as practitioners: the \gls{ngi} by the EC and the \gls{sta} in Germany. The aim is to demonstrate the diversity of designs and implementations of public funding for OSS, demonstrating that a one-size-fits-all approach to measuring the impact of funding is not feasible. We review the \gls{ngi} initiative and the \gls{sta} as examples of public funding because two authors are members of the \gls{ngi} Commons consortium tasked with evaluating the \gls{ngi} initiative\footnote{N.B. While the authors are funded by the EU's Horizon Europe programme under grant agreement number 101135279 (NGI Commons), the EU does not endorse the research outputs of the consortium.} and one author is an employee of the \gls{sta} tasked with evaluating the impact of the \gls{sta}'s funding, thus providing a practitioner perspective on the impact of these funding programmes.

\subsubsection{Next Generation Internet}\label{subsubsec:ngi}
The \gls{ngi} initiative by the EC is a public funding initiative that funds research and development (R\&D) for open internet technologies in the name of an ``Internet of Trust''. As part of Horizon Europe, the \gls{ngi} initiative has provided €140 million for over 1,200 projects between 2019 and 2024, with an additional €32 million allocated for 2024-2027 \cite{ngi_discover_2023}. The majority of funded projects have contributed to OSS projects and/or released solutions in OSS repositories \cite{european_commission_directorate-general_for_communications_networks_content_and_technology_benchmarking_2024}. The \gls{ngi} initiative employs two main funding mechanisms: \gls{rias}, which focus on funding the development of \gls{ngi} technologies, and \gls{csas}, which facilitate scaling efforts through outreach, community building, and international collaboration.  

The \gls{ngi} initiative's \gls{rias} employ cascade funding to distribute funds, allocating 20\% to intermediary coordinators and 80\% to third-party individuals and organisations.\footnote{Cascade funding is also referred to as \gls{fstp}.} The \gls{rias} have employed different funding approaches. First, over €50 million has been distributed in small-to-medium, equity-free grants, ranging from €5,000 to €50,000, to hundreds of recipients, including individuals and start-ups (approximately 800/>1,200 projects by NLNet). These grants are completed by delivery milestones within 12 months, and recipients are paid per milestone completion. According to a \gls{ngi} coordinator, this funding approach ``evaporates the BS factor'' and results in a ``if they fail, they fail fast'' model, which reduces the waste of public funds. Several \gls{rias} have awarded medium-to-large grants, but they have differed in their approach. For example, \gls{ngi} Trust (€7 million) funds the development of privacy and trust-enhancing technologies, using a three-tier funding model ranging from €75,000 for viability studies to €200,000 for commercialisation, with varying matching fund requirements. Meanwhile the European Self-Sovereign Identity Framework Lab (eSSIF-lab) distributed €5.6 million to 56 projects, offering between €15,000 and €106,000 depending on project stage and an additional €155,000 for developing open source self-sovereign identity components. In addition to funding, all NGI0 projects, as well as the NGI0 Review \gls{csa}, have conducted security audits since 2019 on the entire \gls{ngi} portfolio as a systematic measure to improve security and resilience.

A survey of 291 NGI-funded projects by Gartner Consulting offers insights on the impact of the \gls{ngi} initiative  \cite{european_commission_directorate-general_for_communications_networks_content_and_technology_benchmarking_2024}. 63\% of projects have shared their solution in a public OSS repository, and 41\% said their project was part of a larger OSS community effort. 76\% of the funded projects have external communities (other than the core developers and maintainers) contributing to their projects, with 40\% reporting small communities (fewer than 10 people) and 15\%  reporting large communities (over 50 people). Based on the survey results, the authors estimate that each NGI-funded contributor has supported a community of roughly 50 contributors, resulting in a 1:50 multiplier effect, and around 80,000 individuals actively contribute to NGI-funded OSS projects through code contributions, testing, and bug reporting, among others. However, this multiplier effect estimation follows a pragmatically simple calculation.\footnote{The report simply notes ``The multiplier effect regarding the community takes into account an estimation provided by DG CONNECT of the number of  grantees per project: 1.5.''} 

Despite these promising indicators, many challenges hinder the ability to assess the impacts of the \gls{ngi} initiative systematically or at scale. First, the aforementioned estimates are based on a survey of a third of funded projects, leaving us with a partial understanding of the impacts of \gls{ngi} funding. Furthermore, the survey focused on qualitative impacts (e.g. relevance to EU legislation and participation standards bodies) and lacks data required for quantitative analysis (e.g. funding amounts, funding dates, and repository URLs). Beyond this survey, key methodological obstacles include the heterogeneity of funding models and projects, as well as limited data availability. For instance, the \gls{ngi} initiative does not maintain a registry of project-level funding data, such as amounts, dates, and objectives, creating a significant data gap which is difficult to resolve due to the decentralised nature of \gls{ngi}’s funding approach and a lack of systematic record-keeping across \gls{ngi} \gls{rias}. Many grants are given to individuals, which creates difficulties in publishing names and funding due to the General Data Protection Regulation. One \gls{ngi} coordinator described obtaining precise funding data as difficult as ``looking for the hen with the golden eggs.''\footnote{N.B. They used this expression to convey the difficulty of obtaining funding data, rather than to convey greed , as per the idiom.}

\subsubsection{Sovereign Tech Agency}\label{subsubsec:sta}
The \gls{stf} was established in 2022 by the German Federal Ministry for Economic Affairs and Climate Action (BMWK) and hosted by SPRIND GmbH, the Federal Agency for Disruptive Innovation, to strengthen the resilience, security, and sustainability of critical open-source digital infrastructure. The \gls{stf} provides OSS projects with contracts for maintenance and development work starting from €50,000, with no upper funding limit, and supports project durations ranging from 6 to 24 months. In November 2024, the \gls{stf} was reorganised under a newly formed umbrella entity: the Sovereign Tech Agency (\gls{sta}). This transition allowed the \gls{sta} to expand its scope beyond the \gls{stf}’s original focus, establishing a broader array of initiatives and programs. The \gls{stf} continues as a core component within the \gls{sta}, which now operates with an annual budget of approximately €19 million. 

Since its inception, the \gls{stf} has provided €23.5 million in funding to OSS developers and projects for various tasks, including code contributions, security improvements, and infrastructure overhauls. In its pilot phase, the \gls{stf} allocated around €1.3 million to set up the programme and funded nine projects. The pilot included OpenMLS, curl, WireGuard, OpenSSH, Bundler/RubyGems, OpenBGPd, Sequoia-PGP, Fortran Package Manager, and OpenPGP.js/openPGP. A notable example is the investment in the PHP ecosystem, with €205,000 allocated between 2023 and 2024 to support the PHP Foundation. This funding is aimed at addressing outdated infrastructure, enhancing security measures, and improving documentation. Given PHP's foundational role in web development—powering approximately 75\% of websites globally—these efforts were critical for ensuring the sustainability and security of a widely used programming language. Another key investment was in Log4j, a widely-used Java logging library, which received €596,160 in 2023 to improve its security and stability following the critical vulnerabilities exposed during the Log4Shell incident in 2021. This investment supported key enhancements such as fuzz testing, improvements to the project’s release pipeline, and the development of a Software Bill of Materials (SBOM) to better track dependencies and security risks.

In addition to the project-based funding of the \gls{stf}, the \gls{sta} offers a bug resilience programme (Sovereign Tech Resilience), which was launched in early 2023 to enhance the security and resilience of OSS. The resilience programme focuses on reducing technical debt, conducting code security audits, and managing a bug bounty programme to address undiscovered software vulnerabilities proactively. With a budget of €2.5 million over two years, the programme is designed to offer a holistic approach to vulnerability management, ensuring that OSS projects are better prepared to prevent and mitigate security risks. The Sovereign Tech Agency is also piloting a fellowship programme (Sovereign Tech Fellowship) aimed at directly supporting the maintainers of critical OSS projects. This programme aims to pay for the typical work of maintainers up to €78,000 per year per individual. Fellows work across multiple technologies and perform essential tasks such as technical reviews, community management, and security triage.  The fellowship provides both freelance contracting options and a full-time ``maintainer-in-residence'' position, offering maintainers the financial stability to continue their contributions. By investing in the economic stability and security of individuals behind the code, the Fellowships seeks to address structural issues within the OSS community at large, mitigate maintainer burnout, and ensure the long-term sustainability of critical OSS projects. 

\section{A Toolkit for Measuring the Impacts of Public Funding on OSS Development}\label{sec:toolkit}

It is difficult to measure the impacts of funding on OSS development due a number of reasons, including the heterogeneity of OSS projects and ecosystems, the heterogeneity of funding approaches and objectives, and a range of methodological challenges. As a guide through this complexity, we provide a toolkit that may help researchers and practitioners conceptualise and empirically measure the multifaceted impacts of public funding on OSS development. It is informed by methodologies for OSS valuation and community health measurement as well as our first-hand experiences as practitioners tasked with evaluating the impacts of public funding, as discussed in Section~\ref{sec:lit-review}. It is guided by the following motto: If we can meaningfully quantify the impacts of public funding on OSS development, then we can inform funding strategies that are beneficial both to OSS developers and society at large. We underscore that this toolkit is neither prescriptive, nor exhaustive; rather, it is a systematic organisation of key learnings and considerations derived from prior scholarship and our first-hand insights at the \gls{ngi} initiative, the \gls{sta}, and CHAOSS project.

\subsection{Start with Contextual Considerations}\label{subsec:toolkit-context}

\subsubsection{Start with the Funding Objectives}\label{subsubsec:toolkit-objectives}
To state the obvious: not all funding is the same. Different funding instruments---whether milestone-based contracts with deliverables, bug bounty programmes, or general project support---have varying purposes, from innovation to maintenance, and as such varying impacts. Furthermore, one should separate funding at the project and ecosystem levels, as well as funding for project-specific goals and systemic goals, such as infrastructure and developer tools. For example, funding aimed at software development might focus on technological outcomes, such as bug fixes or feature enhancements, while community-oriented funding may focus on contributor diversity or governance changes. 

Understanding the specific objectives of the funding helps to align the impact measurements with the expected outcomes. For example, as described above, both \gls{ngi} and the \gls{sta} fund projects based on a set of milestones that are mutually agreed upon by the funder and the project. As a result, each project works toward different milestones with different goals and objectives, so there is no one-size-fits-all measurement to determine the impacts of these funding programmes. Therefore, one approach is to assess impacts against each project's agreed milestones. However, there is also a case to be made against against such an approach, as examining impacts according to how well the work conforms to specified goals may just be a way to measure compliance rather than impact. In Section~\ref{subsec:toolkit-taxonomy}, we discuss a range of possible economic, social, and technological impacts, which can be both positive and negative, direct and indirect, and manifest over varying time horizons, which provide a broader perspective beyond funding objectives.

\subsubsection{Consider Project Life Stages and Social Structures}\label{subsec:toolkit-maturity}
The impacts of funding vary depending on both a project's life stage and its social structure. For example, a new prototype project will have very different needs and potential outcomes compared to a mature project with an established community. While the former might need funding to build initial features and attract contributors, the latter might require support for security improvements or maintenance work. 

Similarly, considering the social structure of a project is crucial for understanding funding impacts. For instance, using the project types identified by Nadia Eghbal\footnote{As mentioned in Section~\ref{subsubsec:chaoss}), Nadia Eghbal identifies four distinct models of \gls{oss} projects based on their contributor and user ratios: federations, clubs, stadiums, and toys \cite{eghbal_working_2020}.}, funding for a federation-type project  might need to account for complex governance processes and multiple working groups, while funding for a stadium-type project might focus on supporting its small core of maintainers who serve a large user base. These structural differences affect not only how funding should be allocated but also how its impact should be measured.

\subsubsection{Account for Salary Structures and Cost Factors}\label{subsubsec:toolkit-costs}
One should consider salary structures and cost factors across regions and organisations. When similar budgets are allocated to different organisations or regions, they can support different numbers of developers due to varying salary levels. Furthermore, even with access to salary and costs data, its relevance is problematised by the fact that remuneration and the quality of contributions do not always correlate. For example, many highly skilled and experienced OSS developers work in organisations that pay more modest salaries compared to major technology companies, meaning a senior developer in the former may earn less than a junior developer in the latter.

\subsection{A Taxonomy of Economic, Social, and Technological Impacts}\label{subsec:toolkit-taxonomy}

When assessing the impacts of OSS funding, there may be a tendency to focus on technological impacts, which may in part be due to the technological nature of OSS development or the relative ease of measuring technological impacts due to data availability from OSS repositories. However, as discussed in Section~\ref{subsubsec:chaoss}, the potential impacts of funding extend beyond the code itself, and it is crucial to consider economic and social impacts on OSS projects and their wider ecosystems of dependents and users. Furthermore, impacts can be direct or indirect, internal (i.e. affecting the project and its community of contributors) or external (i.e. affecting its ecosystem of dependents or users). It is also important to note that impacts are not linear or unidirectional; funding can lead to both improvements and degradations across different metrics. This multidirectional nature of impacts means that measurement frameworks need to be flexible to capture both positive and negative changes, rather than assuming funding inherently leads to improvements or that metrics only move in one direction. We provide illustrative examples of various impacts in Table~\ref{tab:impact-matrix}, as well as a longer list of question prompts about economic, social, and technological impacts in Appendix~\ref{sec:appendix-impact-questions}. These questions are not exhaustive; rather, they intend to provoke reflection by practitioners about the possible impacts of funding. 

\begin{table}[t]
\centering
\caption{Examples of Social, Economic, Technological Impact Areas of OSS Funding}
\label{tab:impact-matrix}
\small
\setlength{\leftmargini}{1em}
\begin{tabular}{|p{2cm}|p{6.5cm}|p{6.5cm}|}
\hline
& \textbf{Internal Impacts (Project-level)} & \textbf{External Impacts (Ecosystem-level)} \\
\hline
\vspace{1mm}
\textbf{Direct \newline Impacts} & 
\begin{itemize}
    \item \textit{Social:} Contributor retention, community engagement, community events, contributor diversity, work-life balance, reduced burnout, mentorship
    \item \textit{Economic:} Paid developer time / support roles, infrastructure coverage, conference sponsorship,  project-related revenue via various channels
    \item \textit{Technological:} Maintainer responsiveness, commit velocity, code security,  dependency management, documentation quality, consistent releases
\end{itemize}& 
\begin{itemize}
    \item \textit{Social:} User trust, ecosystem community, ecosystem events
    \item \textit{Economic:} Cost savings for adopters, integration/support costs, shared maintenance burden
    \item \textit{Technological:} Stability of APIs, ecosystem-wide security updates, interoperability
\end{itemize} \\
\hline
\vspace{1mm}

\textbf{Indirect \newline Impacts} & 
\begin{itemize}
    \item \textit{Social:} Leadership development, governance and decision-making processes, knowledge preservation, conflict resolution/prevention mechanisms
    \item \textit{Economic:} Job market value for developers, partnership opportunities, academic collaborations, consulting opportunities, funding diversity
    \item \textit{Technological:} Standardisation, interoperability
\end{itemize} & 
\begin{itemize}
    \item \textit{Social:} Cross-project collaboration, training \& education resources, ecosystem engagement
    \item \textit{Economic:} Market growth, job creation, industry cost reduction, start-up creation
    \item \textit{Technological:} Standardisation, research papers, patents, ecosystem-wide security improvements
\end{itemize} \\
\hline
\end{tabular}
\end{table}

\vspace{5em}

\subsection{Temporal Considerations}\label{subsec:toolkit-temporal-considerations}

When measuring the impact of funding on OSS development, it is crucial to consider the effects across different time horizons. The influence of funding can manifest differently in the short, medium, and long term, and these temporal distinctions are essential for a comprehensive understanding of funding impacts.

\begin{enumerate}
    \item \textbf{Short term (<1 year):} These are the immediate effects of funding, which may be easier to observe and attribute to the funding received. Short-term impacts might include immediate increases in development activity, improvements in code quality, or rapid growth in community size.
    \item \textbf{Medium term (1-3 years):} Medium-term impacts may reveal how the project adapts and grows with the resources provided. This period might show the development of new features, expansion of the project's reach, or evolution of the project's governance structures.
    \item \textbf{Long term (>3 years):} Long-term impacts are the enduring effects of funding that manifest over an extended period, which can be the most challenging to measure and attribute directly to funding.
\end{enumerate}

\subsection{Methodological Approaches and Challenges for Impact Measurements}\label{subsec:toolkit-methodological-approaches}
We can employ a range of qualitative, quantitative, or mixed-method approaches, which offer relative strengths and weaknesses, to measure the impacts of public funding on OSS development. We provide a summary of the strengths and weaknesses of these methodological approaches in Table~\ref{tab:methodological-approaches}. This review draws on prior work on empirical methodologies in software engineering research \cite{easterbrook_selecting_2008,runeson_case_2012} and community health methodologies discussed in Section~\ref{subsubsec:chaoss}.

\begin{table}[p]
\centering
\caption{Strengths and Weaknesses of Methodological Approaches to OSS Funding Impact Measurements}
\label{tab:methodological-approaches}
\small
\setlength{\leftmargini}{1em}
\begin{tabular}{|p{3.5cm}|p{3.5cm}|p{3.5cm}|p{3.5cm}|}
\hline
\textbf{Methods \& Data} & \textbf{Strengths} & \textbf{Weaknesses / Challenges} & \textbf{Relevance to OSS Funding} \\
\hline
\vspace{1mm}
\textbf{Qualitative} 
\begin{itemize}
    \item Interviews 
    \item Case studies 
    \item Participant observation 
    \item Focus groups 
    \item Content analysis
\end{itemize} &
\vspace{1mm}
\begin{itemize}
    \item Rich, contextual insights into funding impacts
    \item Uncovers unexpected or indirect impacts
    \item Explores complex interactions between funding and project dynamics
    \item Captures human stories and motivations behind funding impacts
\end{itemize} &
\vspace{1mm}
\begin{itemize}
    \item Limited scalability for large funding initiatives
    \item Potential for researcher and participant bias
    \item Time-consuming and resource-intensive
    \item Difficulty in generalising findings across diverse OSS projects
\end{itemize} &
\vspace{1mm}
\begin{itemize}
    \item Identifying and accessing key stakeholders from OSS projects
    \item Ensuring representative sampling in global, often anonymous contributor bases
    \item Capturing long-term impacts within typical research timeframes
    \item Distinguishing funding impacts from other influencing factors
\end{itemize} \\
\hline
\vspace{1mm}
\textbf{Quantitative} 
\begin{itemize}
    \item Repository mining
    \item Surveys
    \item Econometric modelling
    \item Time series analysis  
    \item Network analysis
\end{itemize} &
\vspace{1mm}
\begin{itemize}
    \item Scalable analysis for large numbers of funded projects
    \item Identifies trends and patterns in funding impacts
    \item Provides quantifiable evidence for policymakers
    \item Leverages digital trace data from repositories
    \item Enables comparative analysis across projects and time periods
\end{itemize} &
\vspace{1mm}
\begin{itemize}
    \item Oversimplifies complex social and technical dynamics
    \item Identifying random variation in funding 
    \item Isolating funding effects from other variables
    \item Data availability/quality
    \item Project workflows affect metric comparability
    \item Difficulty in quantifying intangible outcomes (e.g. security, innovation)
    \item Risk of creating perverse incentives through metric selection/optimisation
\end{itemize} &
\vspace{1mm}
\begin{itemize}
    \item Lack of standardised financial data across OSS projects
    \item Complexity in defining control groups
    \item Accounting for the non-linear and often delayed impacts of funding
    \item Measuring indirect or ecosystem-wide effects of funding
    \item Balancing the need for standardised metrics with project diversity
\end{itemize} \\
\hline
\vspace{1mm}
\textbf{Mixed-Methods} 
\begin{itemize}
    \item Sequential explanatory design (first quantitative, then qualitative)
    \item Sequential exploratory design (first qualitative, then quantitative)
    \item Concurrent triangulation (qualitative \& quantitative simultaneously)
\end{itemize} &
\vspace{1mm}
\begin{itemize}
    \item Balances depth of insight with breadth of analysis
    \item Triangulates findings for increased validity
    \item Adaptable to diverse OSS project contexts
    \item Captures both quantifiable outcomes and complex social dynamics
    \item Allows for iterative refinement of research questions
\end{itemize} &
\vspace{1mm}
\begin{itemize}
    \item Resource-intensive, requiring expertise in multiple methodologies
    \item Complexity in integrating different data types
    \item Potential for conflicting results between methods
    \item Challenges in synthesising diverse data into coherent narratives
    \item Risk of compromising depth or breadth in pursuit of integration
\end{itemize} &
\vspace{1mm}
\begin{itemize}
    \item Aligning qualitative and quantitative data collection in decentralised OSS projects or ecosystems
    \item Balancing the needs of various stakeholders (e.g. developers, policymakers) in research design
    \item Addressing varying timescales of funding impacts across methods
\end{itemize} \\
\hline
\end{tabular}
\end{table}

\subsubsection{Quantitative Approaches}\label{subsubsec:toolkit-methods-quant}
Quantitative methods offer a structured, scaleable approach for measuring relationships between funding and outcomes in OSS projects and their ecosystems. We can differentiate between three main types of quantitative analysis. First, correlational analysis, which is currently the most common approach, helps establish broad patterns and magnitudes of relationships between funding and various outcome measures. Second, quasi-experimental methods attempt to identify causal relationships by isolating quasi-random variation in observational funding data to estimate multiplier effects. Third, experimental approaches, such as randomised controlled trials, aim to establish direct causal relationships through controlled experiments with random assignment of funding; however, such approaches have not been used in OSS funding research to date, in large part due to their impracticality in this context. While each approach offers different levels of causal inference, most existing research relies on correlational analysis, highlighting an opportunity for more rigorous causal investigation through quasi-experimental and experimental methods.

A critical question to consider is which variables to include in the quantitative analysis. Rather than taking a ``big net approach'' to fish for impacts in the wild, we recommend taking a cautious approach to variable selection. As discussed in Section~\ref{subsubsec:toolkit-objectives}, we recommend beginning by accounting for funding objectives. If one seeks to analyse many OSS projects, we suggest considering high-level objectives of a funding initiative rather than the specific milestones agreed between the funder and individual projects. Critical input variables concern funding information, such as allocated funding amounts, the number of funded developers, funding period, and specific objectives (e.g. security audits, feature development, or community building), as they help you know what kinds of relationships to look for and when. 

With regards to output variables, we refer to the taxonomy of potential direct and indirect, internal and external economic, social, and technological impacts in Section~\ref{subsec:toolkit-taxonomy}, as well as a list of relevant variables in Appendix~\ref{sec:appendix-quant-variables}). Platforms like GitHub or GitLab provide access to digital trace data about development activity in OSS projects that can be turned into metrics, such as commits, number of contributors, and forks. Metrics like forks have also been shown to be reliable proxies for usage \cite{vargas_estimating_2024}. Additional data about OSS usage or investment, such as downloads, patents, publications, and dependent projects, may be tracked through third-party sources. Furthermore, variables about various technical and non-technical contributors in a project or its ecosystem \cite{young_which_2021,ferraioli_bringing_2022} and community governance, diversity, and health  \cite{goggins_open_2021,linaker_sustaining_2024} can provide a more comprehensive picture of social dynamics in OSS projects.  

While quantitative methods offer scalable approaches, they are fraught with methodological challenges and limitations. The first challenge concerns data availability. OSS repositories do not publicly report detailed information about funding sources or amounts, making it difficult to assess the exact relationship between funding inputs and project outputs by relying simply on repository data. Beyond funding data, data about OSS usage is infamously scarce \cite{vargas_estimating_2024}, limiting analysis of impact by adoption. Second, a major challenge stems from the non-random nature of funding and the difficulty to isolate funding effects from other influences on project activity (i.e. causal attribution). While an increase in the number of  commits or contributors may be observed during or after a funding period, it is difficult to attribute these changes directly to the funding itself. The non-randomness of funding complicates efforts to create clear causal narratives from quantitative data alone, highlighting the need for mixed-method approaches that can provide both depth and scale in measuring the impacts of funding on OSS projects. Furthermore, significant confounding factors limit our ability to make causal claims about funding and specific outcomes. For example, variables such as the maturity of the project, its pre-existing level of community or contributor activity, or external factors such as varying socio-economic contexts of contributors or market dynamics can all affect project growth and development independent of the funding received or the work contracted. Without accounting for these confounding variables, statistical models run the risk of overestimating or underestimating the impacts of funding.

Additional challenges stem from the different collaboration tools of platforms (``platform affordances'') and development practices across OSS projects, from commit strategies (e.g. micro-commits versus squashed-commits) and code review processes (e.g. direct merge to main branch or code review) to issue management approaches. The same metrics may reflect different underlying activities across projects, complicating cross-project comparisons and large-scale analysis of funding impacts. This necessitates the establishment of project-specific baselines against which to measure changes, rather than comparing absolute metrics across projects. What is more, the measurement of intangible outcomes is difficult. These outcomes are often not captured directly in public data but may instead be inferred from proxies, such as the number of security vulnerabilities fixed, the diversity of contributors, or the creation of new OSS projects. Quantifying these outcomes requires careful selection of metrics and can lead to reliance on external pre-defined measures (e.g. OpenSSF Scorecards). However, these aggregate measures come with the risk of oversimplification, making it challenging to interpret what is being measured (i.e. the estimand of a statistical analysis). 

\subsubsection{Qualitative Approaches}\label{subsubsec:toolkit-methods-qual}
Qualitative methods, such as case studies, semi-structured interviews, and participant observation, are useful strategies for collecting contextual data about the history, role, and impacts of funding on OSS projects. A major strength ofinterviews lies in their ability to provide rich, contextual understandings of a phenomenon by gaining insights into the perspective of the funding recipients, revealing why funding was needed and how it has influenced project dynamics, individual motivations, and community structures \cite{king_interviews_2009,runeson_case_2012,creswell_qualitative_2016}. Furthermore, interviews can uncover the specific ways in which funding has altered a project's trajectory, such as enabling full-time work on previously volunteer-driven projects or facilitating mentorship programmes that were previously out of reach due to limited funds or capacity. One may also learn about unforeseen consequences, such as changes in community dynamics or project governance. Moreover, qualitative methods are useful for capturing intangible outcomes and hard-to-quantify impacts, such as changes in project culture or enhancements in the overall sustainability of the project, which often play a crucial role in the long-term success of OSS projects but can be challenging to measure through quantitative means alone.

However, qualitative methods face significant limitations for measuring OSS funding impacts. One major challenge is stakeholder identification and access. The decentralised nature of many OSS projects can make it difficult to identify and engage key stakeholders for interviews or observations. This is particularly true for projects with large, globally distributed contributor bases, where key individuals may be geographically dispersed or prefer to remain anonymous. Ensuring representativeness is another challenge. It can be difficult to ensure a representative sample in qualitative studies. This limitation can potentially skew findings and make it challenging to draw broader conclusions about funding impacts. Scalability presents another limitation. The time-intensive nature of qualitative research restricts its applicability to large-scale funding initiatives with numerous beneficiaries, such as the \gls{ngi} initiative ($n>1200$). This constraint makes it challenging to draw broad conclusions about funding impacts across diverse projects. Finally, temporal limitations also pose a significant challenge. Capturing long-term impacts of funding may be difficult within typical research time frames, particularly for projects where the effects of funding may take years to fully manifest. This limitation can result in an incomplete picture of funding impacts, potentially missing important long-term effects.

\subsubsection{Mixed-Methods Approaches}\label{subsubsec:toolkit-methods-mixed}
Mixed-method approaches that combine qualitative and quantitative techniques, in theory, offer the best of both words: scalability and contextual depth. Mixed-methods approaches are particularly useful for examining complex OSS ecosystems, where development spans multiple channels and goes beyond repositories. For example, Casari et al. (2023) recommend that researchers collaborate with OSS communities to better contextualise quantitative data \cite{casari_beyond_2023}. Below we discuss three different approaches to sequencing the mixed-methods analysis of OSS funding impacts. \\

\textit{\textbf{Sequential explanatory design (first quantitative, then qualitative)}}\newline

The sequential explanatory design begins with quantitative analysis, followed by qualitative methods to explore the relationship between the funding and these patterns in detail. Such an approach may be particularly useful if one seeks to understand the impact of funding for a large number of projects; for example, it could be applied to assess the impact of the \gls{ngi} initiative’s funding for over 1,200 R\&D projects, many of which contributed to OSS projects and/or released new software in open source repositories. Given the sample size, we require a scalable and systematic approach for collecting and analysing data about these projects, which a quantitative approach is well-suited for.

However, this approach faces challenges. First, we would need to gather repository URLs for NGI-funded projects, and identify which platforms they are hosted on and available options for programmatic data collection (e.g. via APIs). One could use existing survey data, such as the Gartner Consulting survey ($n=291$), then manually fill data holes via internet searches for these repositories. An alternative approach would be to distribute a new survey to all NGI-funded projects, which could collect additional missing data, such as funding amounts, objectives, time period, and project maturity, but it would require coordination with decentralised RIA coordinators and the responsiveness of projects is not guaranteed. Regardless of the approach taken, once we have repository URLs and funding information for some percentage of the projects, we must select the variables for the analysis. Given the diversity of projects in the \gls{ngi} portfolio, it could be sensible to divide projects into buckets based on some criteria and to select variables per bucket. For example, given the range of objectives and project life stages, from proof-of-concepts to feature development in mature projects, one could divide projects into maturity-based buckets. Then, once one has collected data, one may conduct time-series analysis or econometric modelling to identify general trends of changes to activity across all or perhaps among clusters of projects. This analysis may also highlight outliers or unexpected patterns. 

However, given the heterogeneity of the more than 1,200 NGI-funded projects, one should be cautious about using a broad-stroke, quantitative analysis to the many projects, which are diverse by their technology domain, funding approach and objectives, and project life stage, among others. This is where a qualitative analysis of a smaller sample of projects would be useful; for example, by conducting  qualitative case studies on a smaller sample of projects (e.g. $n=10$). Through the analysis of publicly available information (e.g. documentation, blogs, mailing lists, etc) and interviews with the funding recipients, one can gain a more nuanced understanding of how funding was used by and impacted a project. This qualitative phase also allows us to test the usefulness of our quantitative findings or lack thereof, and uncover impacts that were not captured by the quantitative analysis. \\

\textit{\textbf{Sequential exploratory design (first qualitative, then quantitative)}} \newline

Sequential exploratory design begins with qualitative analysis, followed by quantitative analysis. Consider a scenario where we want to evaluat the impacts of the \gls{sta}’s pilot phase funding on a smaller number of OSS projects ($n=9$). We may begin with a thorough review of funding agreements for all projects in the \gls{sta} pilot phase, providing a contextual understanding of what the funding was intended to achieve. Next, we conduct semi-structured, exploratory interviews with key stakeholders in each funded project. These interviews allow us to gather rich, contextual data about the funding experience. Questions might include: ``How did you utilise X amount of funding to achieve Y objective?'', ``What specific activities or outputs were facilitated by this funding?'', ``Are there digital traces of these activities that could be programmatically scraped and quantitatively analysed, and if so on which platforms?'', and ``In your opinion, what variables should we control for in our analysis?'' 

This qualitative phase offers several advantages. It allows for the discovery of unexpected impacts that may not have been captured in quantitative metrics. It also provides insight into the human experiences and decision-making processes behind the numbers. However, it is time-intensive and may be subject to individual biases or selective recall. Using these qualitative insights, we then inform our subsequent quantitative analysis. We can identify relevant data sources, mine data from repositories with a focus on areas highlighted in interviews, and conduct targeted quantitative analyses that account for project-specific contexts and potential confounding variables. This approach ensures that our quantitative analyses are grounded in the realities and nuances of each project's funding objective and experience. It allows for a more targeted and meaningful quantitative analysis, potentially uncovering patterns or relationships that might have been overlooked in a purely quantitative approach. However, this method may be less suitable for large-scale evaluations due to the time-intensive nature of the qualitative phase. It also requires careful consideration to ensure that insights from the qualitative phase are appropriately translated into quantitative measures. \\

\textit{\textbf{Concurrent triangulation approaches (e.g. maturity models)}} \newline

Maturity models are a concurrent triangulation approach that combine qualitative assessments with quantitative metrics at the same time to create a holistic view of the development of an OSS project \cite{becker_developing_2009,poppelbus_what_2011}. A maturity model approach might involve defining relevant dimensions of project maturity (e.g. code quality, community size, documentation completeness), establishing qualitative criteria and quantitative metrics for each dimension, collecting data through surveys, interviews, and automated repository analysis, and analysing changes in maturity levels in relation to funding inputs. For example, maturity models have been proven to be an effective approach to understand the health of young and data-scarce OSS ecosystem health \cite{van_vulpen_health_2017}. This approach offers a standardised yet flexible method for comparing diverse projects and allows for tracking changes over time, potentially correlating these with funding inputs. However, developing a comprehensive and relevant maturity model is time-consuming, and there may be disagreement among stakeholders about what constitutes ``maturity'' in different dimensions.

\subsection{Options for and Hazards of Multiplier Effect Estimations}\label{subsec:toolkit-multiplier-effects}

With this knowledge, let us now consider how we might estimate multiplier effects of OSS funding as one form of  quantitative evidence of funding impacts. 

\subsubsection{Conceptualising Multiplier Effects of OSS Funding}\label{subsubsec:toolkit-ultiplier-effect-concept}
In the context of R\&D,  the economic multiplier is the measurement of how much additional economic value is generated for each euro spent on R\&D \cite{griliches_issues_1979}.  It is also relevant to consider knowledge spillovers (i.e. when R\&D conducted by one entity creates value for other entities without compensation to the innovator) and social rates of return (i.e. the total benefits to society from R\&D investment, including both private returns to the innovator and spillover benefits to others). Multiplier effects are often measured through econometric analyses that quantify the relationship between an investment and direct and indirect outcomes over time \cite{griliches_search_1992,hall_measuring_2010}. Relevant variables include input measures (e.g. expenditure on salaries and equipment), output measures (e.g. patents, publications, or products), time lags (i.e. accounting for delays between investment and observable outcomes), and spillover effects (i.e. impacts on related industries).

The multiplier effect of OSS funding may be understood as the generation of value that exceeds the initial funding amount or objective. This may include community contributions, technological innovations, community or ecosystem growth, or downstream economic benefits. This calculation requires the adaptation of methodologies to the collaborative nature of OSS development, where value creation and capture operate differently than in the typical R\&D processes and structures found in labs. Specifically, contributors have the freedom to join or leave projects at will, development paths can fork into various directions, and improvements made by one contributor are instantly accessible to everyone. These dynamics can amplify positive multiplier effects through increased collaboration and knowledge sharing, but they can also lead to negative multiplier effects, such as by creating unsustainable dependencies.

\subsubsection{Measuring Multiplier Effects of OSS Funding}\label{subsubsec:toolkit-ultiplier-effect-measure}
There is a serious risk of oversimplifying complex dynamics by reducing them to single metrics, and as such it is important to measure multiplier effects with caution and with an awareness of the limitations of the data and methods used in such measurements. Due to the quantitative nature of such measurements, the concerns are similar to those elaborated in Section~\ref{subsubsec:toolkit-methods-quant}. Above all, it is challenging to account for the non-random nature of funding and to  isolate funding effects from other influences on project activity, as projects that receive funding often differ systematically from those that do not in ways that affect their outcomes. For instance, more mature or popular projects might be more likely to receive funding and also more likely to grow independently of funding. This selection bias makes it difficult to establish causal relationships between funding and observed outcomes. What is more, it is difficult to quantify intangible benefits, and time lags between funding periods and observable impacts can exceed typical measurement periods, making it difficult to capture the full scope of impacts. There are also important normative or behavioural considerations: the choice of multiplier effect variables itself might create perverse incentives, leading projects to optimise for measured outcomes at the expense of factors that are important to the project and its community.

For these reasons, measuring multiplier effects of OSS funding requires careful consideration of the following factors:

\begin{itemize}
    \item The bidirectional nature of funding impacts, including both positive and negative multiplier effects
    \item The complex network effects and externalities in OSS ecosystems
    \item The role of community dynamics and governance in mediating funding impacts
    \item The potential for funding to affect project independence and sustainability
    \item The distribution of benefits across different stakeholder groups (e.g. developers, dependents, and users)
\end{itemize}

Given these considerations, estimating the multiplier effect of OSS funding should be approached with care and ideally combine quantitative and qualitative metrics that provide a comprehensive, contextual understanding of funding in OSS projects. To aid such measurements, we provide a list of relevant input, output, and control variables in Appendix~\ref{sec:appendix-quant-variables}. Input variables include the funding amount and the number of funded developers; while output variables may be technological (e.g. maintainer responsiveness or growth in dependent projects), economic (e.g. project revenue or cost savings for adopters), or social (e.g. contributor diversity or community events). As noted in Section~\ref{subsec:toolkit-taxonomy}, these economic, social, and technological outputs can be direct or indirect as well as internal or external to the project. Relevant control variables include project life stage, its pre-existing community size, and concurrent market trends. 

\section{Discussion}\label{sec:discussion}
\subsection{Key Components of the Toolkit }\label{subsec:discussion-toolkit}

The provision of public funding is increasingly considered important as a lever for supporting the maintenance, security, and long-term sustainability of OSS projects that form a critical part of our digital infrastructure, yet we still have a limited understanding of the impact of funding and a lack of consensus on how to meaningfully assess or measure impact. We have sought to address this gap by reviewing prior scholarship and documenting our first-hand learnings in the form of a practical toolkit that may inform attempts to measure the impacts of public funding on OSS development. 

Core to this discussion is the following question: How do we capture the myriad ways in which public funding can shape---positively or negatively, directly or indirectly---the development, sustainability, and evolution of OSS projects and their wider ecosystems? While the diversity of OSS projects and ecosystems complicates efforts to develop standardised approaches to impact measurement, our toolkit seeks to offer some practical ways forward. For example, it is important to start with the objectives of the funding and consider OSS projects as socio-technical systems. If one is the funder, one should have a record of the funding objectives, and one should design one’s impact measurement with those objectives in mind. It is important to look beyond the code; that is, to consider the various social and economic impacts of funding OSS development. The toolkit provides guidance on the various potential impacts (see Section~\ref{subsec:toolkit-taxonomy}) as well as methodological options for and hazards of estimating multiplier effects of OSS funding see Section~\ref{subsec:toolkit-multiplier-effects}). 

We discussed the merits and limitations of qualitative and quantitative methods. While quantitative measurements of \gls{roi} or multiplier effects may be desired by decision-makers to justify public spending and optimise funding strategies, we caution against hasty calculations that do not consider the methodological challenges and hazards involved in quantitative measurements. Qualitative methods, in turn, may yield rich and contextual insights into the role and impact of funding within OSS projects, but do not scale and do not produce desired quantitative evidence. Mixed-method strategies may balance these trade-offs, though our examples highlighted that they face challenges, too.

\subsection{Broader Questions and Challenges}\label{subsec:discussion-challenges}
We note that measuring OSS project impacts through metrics carries inherent risks and normative implications. Indicators often become interventions themselves, potentially steering project development in unintended directions \cite{davis_indicators_2012}.\footnote{We recommend reading ``Indicators as Interventions: Pitfalls and Prospects in Supporting Development Initiatives'' by Davis and Kingsbury (2012), especially the ``Indicators: Pathologies and Pitfalls'' table on page 3, for a thorough discussion of the normative implications and potential risks of creating quantitative indicators \cite{davis_indicators_2012}.} The diverse values and priorities of OSS developers must be carefully translated into meaningful metrics for funders. While multiplier effects provide valuable insights, they risk oversimplifying complexity and creating misaligned incentives. A balanced approach combining qualitative and quantitative measurements may help preserve nuance while providing necessary accountability.

The impacts of funding cessation also warrant careful consideration. Research suggests that funded projects often face sustainability challenges post-funding, as evidenced by the Prototype Fund's evaluation finding that over half of their funded projects struggled to secure follow-up funding \cite{galati_evaluation_2023}. This raises critical questions about project sustainability: Does initial funding create dependencies that make projects more vulnerable when funding ends? How can we better prepare projects for post-funding sustainability? More research is needed to understand these transition periods and develop strategies to support projects through funding changes.

A further question to consider is how responsive this toolkit is to various funders in the public sector and beyond, from companies to philanthropic foundations. While the objectives of a company may differ from a government (e.g. focus on specific business needs), the toolkit's recommendations to account for funding objectives, cost factors, and various economic, social, and technological impacts should remain relevant and responsive to companies. We seek to test this hypothesis with diverse funders through future workshops, research, and iterations of this toolkit.

\subsection{Future Work}\label{subsec:discussion-future}
Current understanding of OSS funding impacts remains incomplete in several key areas. Significant knowledge gaps exist in our understanding of long-term effects, the effectiveness of different funding models, and potential unintended consequences, such as impacts on volunteer motivations or creating dependencies. These gaps are particularly concerning given the importance of OSS as digital infrastructure and the challenges projects face, including maintainer burnout and security vulnerabilities. Further research is needed to examine the long-term sustainability of funded projects, compare funding models, and develop adaptable metrics for assessing impact across diverse contexts. Understanding how funding influences the broader OSS ecosystem, including its effects on innovation and collaboration, is also crucial. Addressing these gaps requires collaboration among the OSS developer community, funders, policymakers, and researchers to develop a deeper understanding of the multifaceted impacts of various funding approaches. Such efforts are vital to ensure that funding initiatives support, rather than undermine, OSS developers.

\section{Conclusion}\label{sec:conclusion}
This paper presented a toolkit for measuring the impacts of public funding on OSS development. The toolkit is not prescriptive, nor is it exhaustive; rather, we provide practical recommendations and discuss potential hazards for funding impact measurements. It starts with a discussion of key contextual considerations to inform impact measurements, such as accounting for funding objectives, project life stages and social structures, and  regional and organisational cost factors. Then, we present a taxonomy of social, economic, and technological impacts, which can be both positive and negative, direct and indirect, internal (i.e. within a project) and external (i.e. among a project's ecosystem of dependents and users), and manifest over various time horizons. Then, we provide guidance for the use of quantitative, qualitative, and mixed-methods approaches to measure such impacts, as well as the measurement of multiplier effects of funding. While multiplier effect metrics are useful for decision-makers who require quantitative evidence about returns on investment to justify public spending and optimise funding strategies, we highlight a number of methodological challenges, from data availability to the non-randomness of funding, that complicate making meaningful quantitative estimates. This toolkit is not exhaustive and we still have major gaps in our understanding of the impacts of funding and optimal impact measurements. We, therefore, call for diverse stakeholders---OSS contributors (technical and non-technical), policymakers, funders, researchers, and others---to join this effort to develop frameworks for meaningful, evidence-based evaluations of the multifaceted impacts of public funding on OSS development.

\newpage

\bibliographystyle{ieeetr}  
\bibliography{references}

\section{Acknowledgements}
We would like to thank the following individuals for their advice and/or feedback: Marco Berlinguer, Michael Brennan, Jean-Luc Dorel, Ryan Ellis, Nathalia Foditsch, Matthew Gaughan, Jeff Gortmaker, Bastien Guerry, Karolina Gyurovszka, Valérian Guillier, Anna Hermensen, Benedict Kingsbury, Fiona Krakenbürger, Michiel Leenaars, Johan Linåker, Georg Link, Katharina Meyer, Samtag Prakke, Mirko Presser, Fabien Rezac, Lorenzo Sciandra, Powen Shiah, Katherine Skinner, Paul van Vulpen, Chrys Wu, as well as the members of the \gls{ngi} Commmons consortium.

\section{Funding Declaration}
Cailean Osborne and Mirko Boehm are funded by the EU's Horizon Europe programme under grant agreement number 101135279 (NGI Commons). Dawn Foster is funded by an Alfred P. Sloan Foundation grant for the CHAOSS project. Paul Sharratt is employed and funded by the Sovereign Tech Agency. The views and opinions expressed are those of the authors only and do not necessarily reflect those of the funders.

\newpage
\appendix
\section{Prompt Questions about Potential Technological, Economic, and Social Impacts}\label{sec:appendix-impact-questions}
\subsection{Technological Impacts}\label{subsubsec:toolkit-technological-impacts}

\textbf{Direct impacts:}
\begin{itemize}
    \item How many critical bugs or security vulnerabilities were resolved due to the funding?
    \item Has the funding improved maintainer responsiveness to issues and pull requests? And how has this affected the project's development velocity and community engagement?
    \item  What measurable improvements in code quality metrics (e.g. test coverage, cyclomatic complexity, documentation completeness) have been achieved post-funding?
    \item How have user experience (UX) metrics and user feedback evolved since receiving funding?
    \item What improvements in documentation, tutorials, and user support resources have been achieved with the funding?
    \item  How has the project's technical debt been reduced as a result of the funding, and what impact has this had on development activity?
    \item  What new technologies or development methodologies has the funding enabled the project to adopt, and how have these improved the project's capabilities?
    \item Has the funding led to rushed development of new features, potentially introducing new security vulnerabilities or leading to the neglect of other areas of the project?
\end{itemize}

\textbf{Indirect impacts:}
\begin{itemize}
    \item How has the funded project contributed to the establishment or evolution of industry standards or best practices, and what is the scope of their adoption?
    \item How has the funding influenced collaborations with other OSS projects or organisations?
\end{itemize}

\subsection{Economic Impacts}\label{subsubsec:toolkit-economic-impacts}

\textbf{Direct impacts:}
\begin{itemize}
    \item How many full-time equivalent positions has the funding created or sustained, and how does this compare to pre-funding levels?
    \item What percentage increase in project-related revenue can be directly attributed to development enabled by the funding, and through which channels (e.g. donations, commercial contracts, partnerships, services, etc.)?
    \item What quantifiable cost efficiencies (e.g. in development time, infrastructure costs) have been realised due to the funding, and how do these translate to financial savings?
    \item Has the funding created an over-reliance on a single funding source, potentially compromising the project's long-term sustainability?
\end{itemize}

\textbf{Indirect impacts:}
\begin{itemize}
    \item How has the funding influenced the project's contribution to innovation in its sector, as measured by research publications, standards, patents, new features, or other?
    \item How has the funding impacted the project's ability to compete with proprietary alternatives, and what are the associated economic implications?
    \item How has funding impacted trust from enterprise users, and how has this affected their security auditing costs?
    \item What critical business operations or infrastructures now depend on the funded project?
    \item How has funding contributed to skill development and career advancement in the broader tech workforce?
    \item What impact has the project had on reducing software licensing costs at a societal level?
    \item What spillover effects has the funding had on the broader OSS ecosystem, such as the creation of complementary OSS projects or commercial products or services?
    \item What is the estimated economic impact of productivity gains for businesses using the funded OSS project?
    \item Has the project's growth due to the funding led to increased maintenance costs that may not be sustainable in the long term?
\end{itemize}

\subsection{Social Impacts}\label{subsubsec:toolkit-social-impacts}

\textbf{Direct impacts:}
\begin{itemize}
    \item How has the funding affected the diversity (e.g. gender, geographical, skills) of new contributors, and how does this compare to pre-funding demographics?
    \item What measurable impact has the funding had on the project's mentorship programs or onboarding processes for new contributors?
    \item How has the absolute number and retention rate of contributors changed post-funding?
    \item What impact has funding had on the project's ability to support and retain existing contributors?
    \item How has the funding influenced the project's ability to organise or participate in community events and conferences, and what has been the reach of these activities?
    \item Has the introduction of funding created tensions or conflicts within the community, particularly between funded and unfunded contributors?
    \item Has the funding led to a shift in project governance that excludes or marginalises certain community members?
\end{itemize}

\textbf{Indirect impacts:}
\begin{itemize}
    \item What changes in community engagement metrics (e.g. mailing list activity, issue response times, pull request review times) can be observed post-funding?
    \item Has the project's growth due to funding led to a loss of community culture or a shift in culture that alienates contributors?
    \item What societal benefits have emerged from increased accessibility to the software (e.g., digital inclusion, reduced costs for public institutions)?
    \item How has funding influenced the project's impact on digital sovereignty and technological independence?
    \item Has the increased professionalisation of the project due to funding created barriers to entry for new or casual contributors?
\end{itemize}

\newpage
\section{Variables for Quantitative Measurements of OSS Funding Impacts}\label{sec:appendix-quant-variables}

\subsection{Dependent Variables}

\subsubsection{Technological Impact}
\begin{itemize}
    \item \# pull requests, issues, and commits
    \item \# lines of code added or modified
    \item Time to first response by maintainer
    \item Time to close issue
    \item \# releases or new features
    \item \# users across installations
    \item \# downloads or installations
    \item \# forks and clones
    \item Code quality metrics (e.g. maintainability, test coverage)
    \item \# patents filed
    \item \# new projects initiated
    \item \# standards organisations collaborated with
\end{itemize}

\subsubsection{Economic Impact}
\begin{itemize}
    \item \# jobs supported (e.g. \# full-time employees)
    \item Revenue generated by the project or company (in €)
    \item Amount of follow-on funding or investment (in €)
    \item Cost savings for organisations using the project
    \item \# commercial contracts or partnerships
    \item Market value of the project or company (if applicable)
    \item Economic value of volunteer contributions
    \item \# spin-off projects or companies
\end{itemize}

\subsubsection{Social Impact}
\begin{itemize}
    \item Growth in \# contributors
    \item \# active maintainers
    \item Contributor absence factor 
    \item Bus factor before and after funding period
    \item Growth in \# users
    \item Metrics for diversity, equity, and inclusion (e.g. CHAOSS metrics)
    \item Growth in stars, forks, and watchers on GitHub
    \item Geographic distribution of contributors
    \item \# new contributors mentored
    \item \# educational initiatives or workshops conducted
    \item Community engagement metrics (e.g. forum activity, meetups)
    \item \# academic collaborations or citations
    \item Impact on skills development in the community
    \item \# speaking engagements or conference presentations by project members
\end{itemize}

\subsection{Independent Variables}
\begin{itemize}
    \item Funding objective(s)
    \item Funding amount
    \item Funding duration or distribution schedule

\end{itemize}

\subsection{Control Variables}

\subsubsection{Project Characteristics}
\begin{itemize}
    \item Project age at time of funding
    \item Project life stage (e.g. inception, maturity) 
    \item Experience level of project leaders
    \item Changes in project leadership
    \item Software category (e.g. library, protocol, devtools, application, decentralised platform)
    \item License type
    \item Governance model (e.g. BDFL, committee, foundation)
    \item Previous funding history and/or other funding sources
\end{itemize}

\subsubsection{Community Factors}
\begin{itemize}
    \item Initial community size before funding
    \item Contributor absence factor 
    \item Bus factor before and after funding period
    \item Presence of corporate backers
    \item Geographic concentration of core contributors
    \item Proportion of volunteer vs. paid contributors
\end{itemize}

\subsubsection{Technological Factors}
\begin{itemize}
    \item Programming language 
    \item \# outdated dependencies 
    \item Architectural complexity
    \item Dependency on other projects or technologies
\end{itemize}

\subsubsection{Market and Industry Factors}
\begin{itemize}
    \item Market size for the software category
    \item Level of competition (both open source and proprietary)
    \item Industry trends and technological shifts
    \item Regulatory environment
\end{itemize}

\subsubsection{External Factors}
\begin{itemize}
    \item Economic conditions (e.g. recession)
    \item Major world events (e.g. COVID-19 pandemic)
    \item Major technological events (e.g. the emergence of AI co-pilots)
    \item Changes in related technologies or standards
\end{itemize}
    
\newpage

\end{document}